\documentclass[12pt]{article}
\usepackage{graphicx}
\usepackage{latexsym}
\usepackage{wrapfig}
\usepackage{url}
\usepackage{here}

\let\oldenumerate\enumerate
\renewcommand{\enumerate}{
   \oldenumerate
   \setlength{\itemsep}{1pt}
   \setlength{\parskip}{0pt}
   \setlength{\parsep}{0pt}
}
\let\olditemize\itemize
\renewcommand{\itemize}{
   \olditemize
   \setlength{\itemsep}{1pt}
   \setlength{\parskip}{0pt}
   \setlength{\parsep}{0pt}
}

\newcommand\pubdate{\today}

\def\univ{Department of Particle \& Nuclear Physics, \\School of High Energy Accelerator Science, \\The Graduate University for Advanced Studies (SOKENDAI), Ibaraki, Japan}
\def\co-univ{High Energy Accelerator Research Organization (KEK), Ibaraki, Japan}

\def\Title#1{\begin{center} {\Large #1 } \end{center}}
\def\Author#1{\begin{center}{ \sc #1} \end{center}}
\def\Address#1{\begin{center}{ \it #1} \end{center}}

\newcommand\pubblock{\rightline{\begin{tabular}{l} \pubdate  \end{tabular}}}
\newenvironment{Abstract}{\begin{quotation}  }{\end{quotation}}
\newenvironment{Presented}{\begin{quotation} \begin{center} 
             PRESENTED AT\end{center}\bigskip 
      \begin{center}\begin{large}}{\end{large}\end{center} \end{quotation}}
\def\Acknowledgements{\bigskip  \bigskip \begin{center} \begin{large}
             \bf ACKNOWLEDGEMENTS \end{large}\end{center}}

\def\beq{\begin{equation}}
\def\eeq#1{\label{#1}\end{equation}}
\def\eeqn{\end{equation}}

\def\beqa{\begin{eqnarray}}
\def\eeqa#1{\label{#1}\end{eqnarray}}
\def\eeqan{\end{eqnarray}}

\let\bar=\overbar

\def\Dslash{\not{\hbox{\kern-4pt $D$}}}
\def\dslash{\not{\hbox{\kern-2pt $\del$}}}

\def\msb{{\bar{\ssstyle M \kern -1pt S}}}

\begin{document}
\begin{titlepage}
\pubblock

\vfill
\Title{DAQ Development for Silicon-On-Insulator Pixel detectors}
\vfill
\Author{
 Ryutaro Nishimura${}^{1}$,
 Yasuo Arai${}^{2}$,
 Toshinobu Miyoshi${}^{2}$ 
 }
\Address{
 ${}^{1}$\univ \\
 ${}^{2}$\co-univ}
\vfill
\begin{Abstract}
We are developing DAQ for Si-pixel detectors by using a Slicon-On-Insulator (SOI) technology. 
This DAQ consists of firmware works on SEABAS (Soi EvAluation BoArd with Sitcp) DAQ board and software works on PC. 
We have been working on the development of firmware/software. Now we accomplished to speed up the readout ($\sim$ 90Hz) and to add a function for frame rate control.
This is the report of our development work for the High Speed DAQ system.
\end{Abstract}
\vfill
\begin{Presented}
International Workshop on SOI Pixel Detector (SOIPIX2015), \\
Tohoku University, Sendai, Japan, 3-6, June, 2015.
\end{Presented}
\vfill
\end{titlepage}
\def\thefootnote{\fnsymbol{footnote}}
\setcounter{footnote}{0}

\section{Introduction}

\subsection{SOI Detector}

\begin{figure}[H]
\centering
\includegraphics[height=6cm]{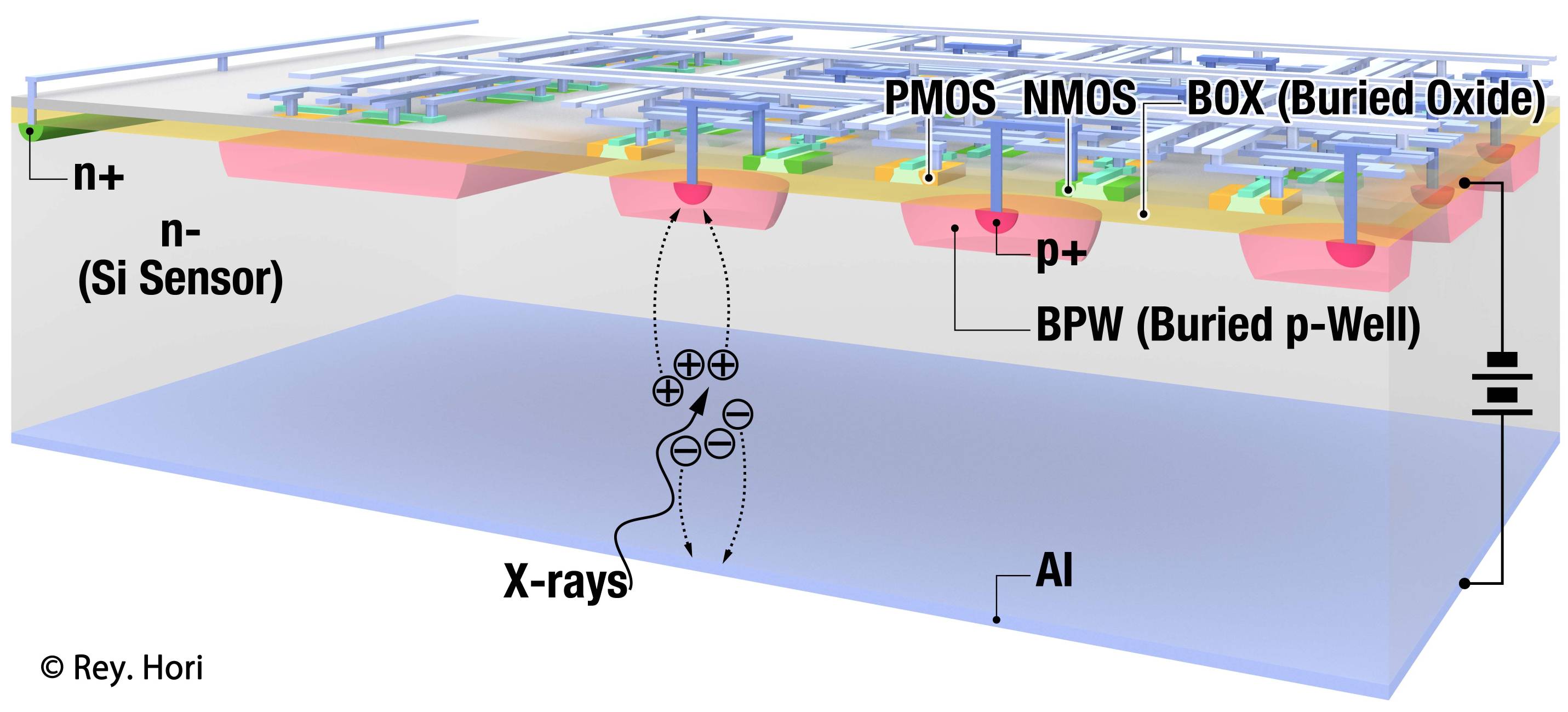}
\caption{The structure of SOI Detector.}
\label{fig:soistruct}
\end{figure}

\noindent
SOI Pixel detectors are being developed by a SOIPIX collaboration led by KEK. 
They are based on a 0.2 um CMOS fully-depleted (FD-) SOI process of Lapis Semiconductor Co., Ltd \cite{soi}. 
A detector's structure image is shown in Fig \ref{fig:soistruct}. 
SOI detector consists of a thick and high-resistivity Si substrate for sensing part, and a thin Si layer for CMOS circuits \cite{soi}. 
An SOI detector has no bump bonding, therefore the application has low capacitance, low noise, high gain, and low material budget. 
It can run fast with low power. 

\begin{wrapfigure}{l}{6cm}
\centering
\includegraphics[height=4cm]{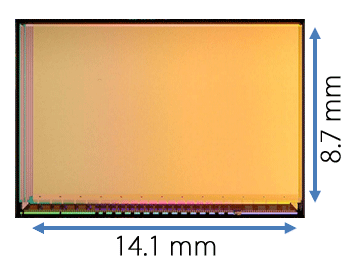}
\caption{The photo of INTPIX4.}
\label{fig:intpix4}
\end{wrapfigure}

For Examination of DAQ system, used the integration type SOI pixel detectors, named INTPIX4 \cite{soi2}. 
The pixel size is $17 \mu m$ squares, a number of pixels are $832 \times 512$, and a sensitive area is $14.1 \times 8.7 mm^2$. 
This detector consists of 13 blocks ($64 \times 512$ pixels / block) and each blocks has independent channels of analog output for parallel readout.
Photo of INTPIX4 is shown in Fig \ref{fig:intpix4}.

\clearpage

\subsection{SEABAS DAQ system}

\begin{figure}[H]
\begin{tabular}{c}
\begin{minipage}{0.45\hsize}
\centering
\includegraphics[width=6cm]{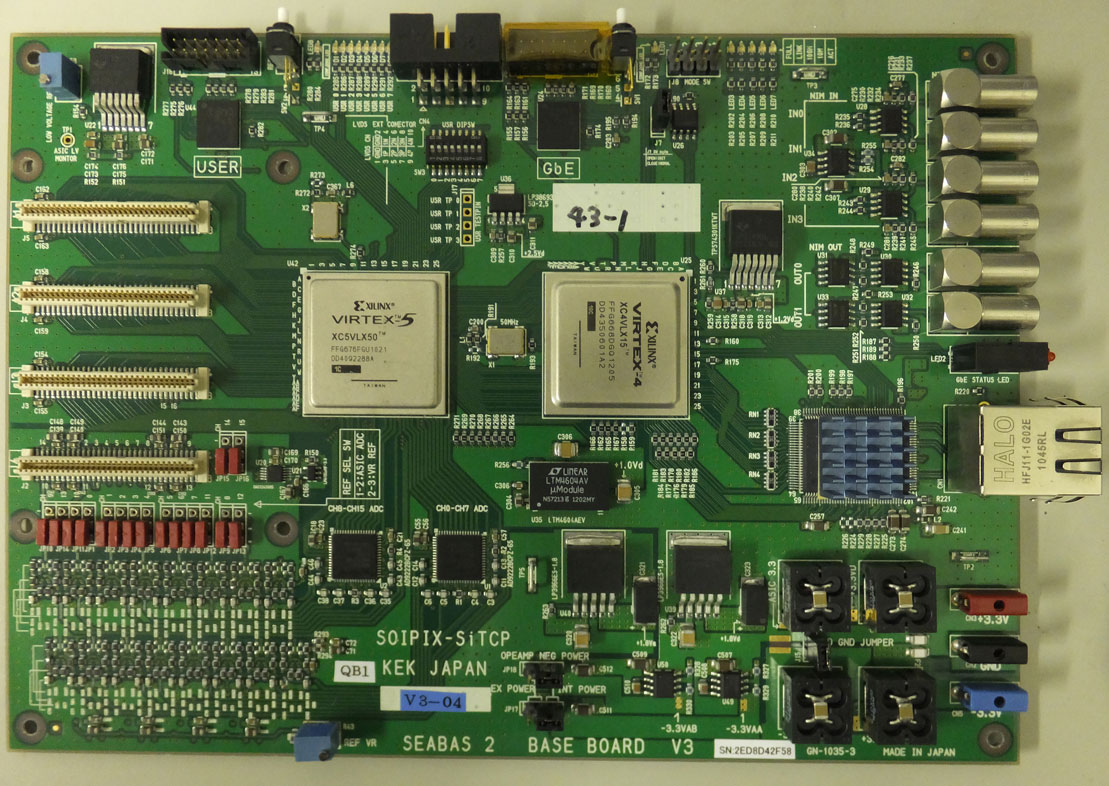}
\caption{The photo of SEABAS2.}
\label{fig:seabas2}
\end{minipage}
\begin{minipage}{0.55\hsize}
\centering
\includegraphics[width=7cm]{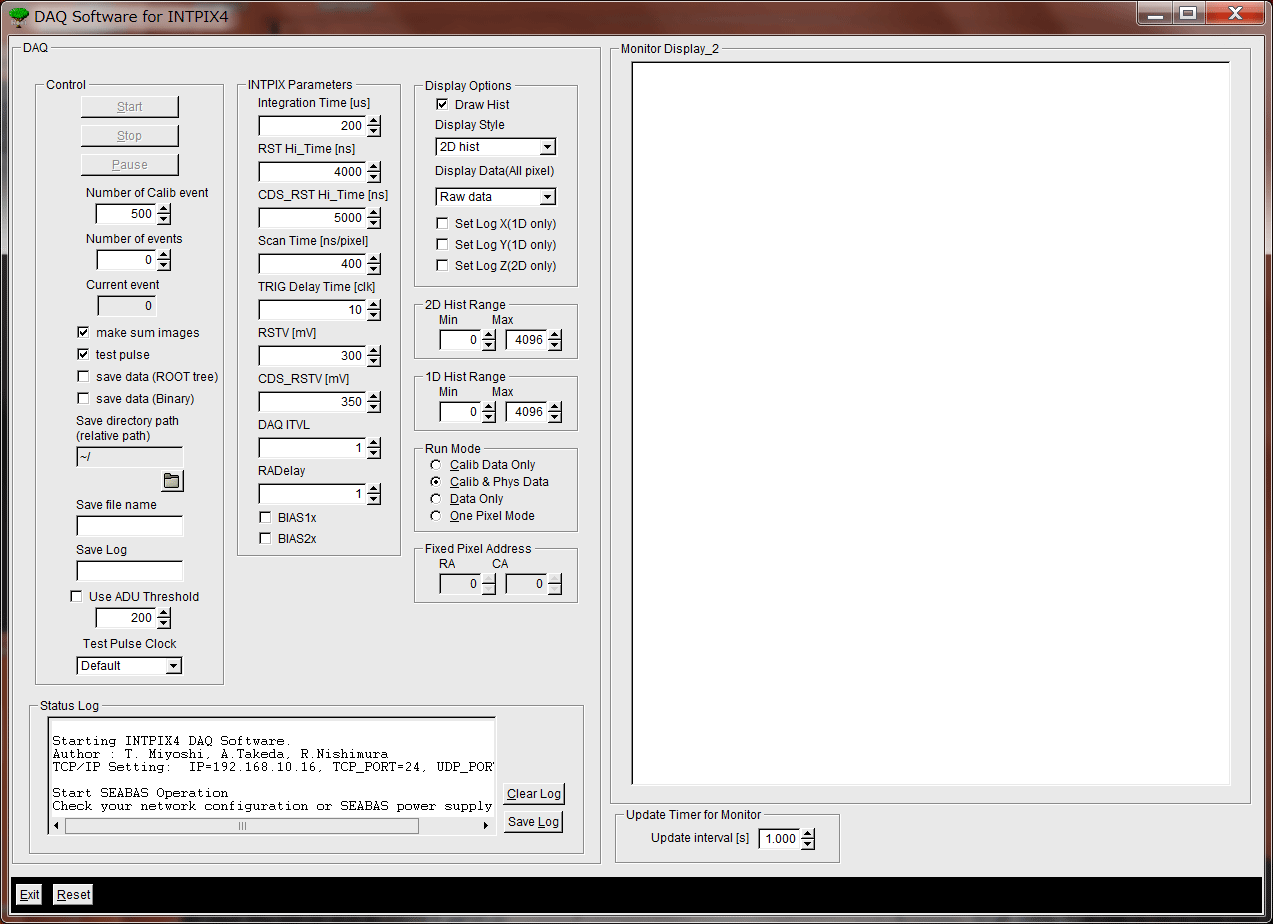}
\caption{The working image of DAQ Software.}
\label{fig:olddaq}
\end{minipage}
\end{tabular}
\end{figure}

\begin{figure}[H]
\centering
\includegraphics[width=14cm]{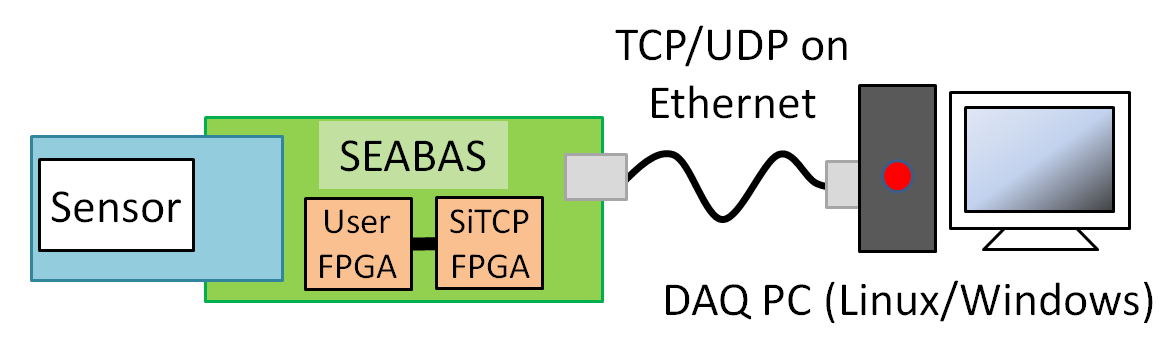}
\caption{The schema of SEABAS DAQ System.}
\label{fig:daqsys}
\end{figure}

\noindent
INTPIX4 is read out through the board called SEABAS 2 (Soi EvAluation BoArd with Sitcp 2 \cite{seabas}).
SEABAS 2 is the 2nd Generation of SEABAS board.
Fig \ref{fig:seabas2} is the photo of SEABAS 2.

This board has FPGAs for Gigabit Ethernet and a user circuit, and 12-bit ADC for convert detector's analog output. 
When SEABAS2 transfer INTPIX4's output after ADC conversion, 
total data size is 6,815,744 bit (ADC 12bit + padding 4bit = 16 bit per pixel) per frame. 
DAQ system consists of firmware work on SEABAS2 user circuit FPGA, and software work on PC (shown in Fig \ref{fig:olddaq}). 
Between PC and SEABAS2 are connected by Gigabit Ethernet and communicate TCP/UDP protocol.
The schema of SEABAS DAQ is shown in Fig \ref{fig:daqsys}.

\clearpage

\subsection{Needs of High Speed DAQ}

\begin{wrapfigure}{l}{6cm}
\centering
\includegraphics[height=1.5cm]{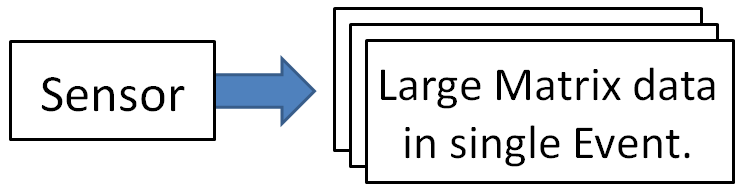}
\caption{The schema of Readout from sensor.}
\label{fig:nhs}
\end{wrapfigure}

SOI Pixel sensor outputs a large quantity of data at a time. 
Fig \ref{fig:nhs} is a simple schema of readout from sensor.
For example, 6,815,744 bit/frame in the case of INTPIX4. 
Therefore, fast data transfer is important for quick and high resolution measurement.
And if use DAQ for temporal response measurement, stability of transfer rate is also important.

\section{Methods}

Our approach for high speed \& stable transfer DAQ is total 3 points in software and firmware.

\subsection{Abstraction and Hierarchization}

\begin{figure}[H]
\centering
\includegraphics[width=14cm]{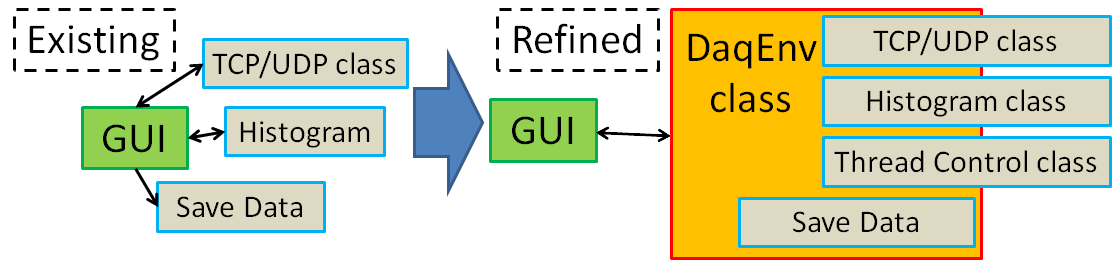}
\caption{The Schema of Abstraction and Hierarchization.}
\label{fig:method_ah}
\end{figure}

\noindent
``Abstraction and Hierarchization" is the one of the refine approach in software region. 
The schema of this refine is shown in Fig \ref{fig:method_ah}.

In existing system, all functions are controlled from GUI directly. 
And many function's codes are written without classified. 
This situation causes confusion in software development, such as "Spaghetti Code". 

Therefore, we refined this software structure. 
We separated function's codes without directly related to GUI, 
and function's codes were classified and concentrated to some classes and functions under DaqEnv class. 
All classes completes in its own as much as possible. 
DaqEnv is the abstraction class for absorbing all environment difference. 
GUI access every functions via DaqEnv class. 

This refine gives 2 merits.

\begin{itemize}
 \item Easy to add/delete function. \\
       We only have to fix related class's code. Other codes are needless to touch. 
 \item Easy to apply for any SOI Sensors. \\
       To apply this software for another sensor, only few codes need modify. 
\end{itemize}

\subsection{Multi Thread (MT) Processing}

\begin{figure}[H]
\centering
\includegraphics[width=14cm]{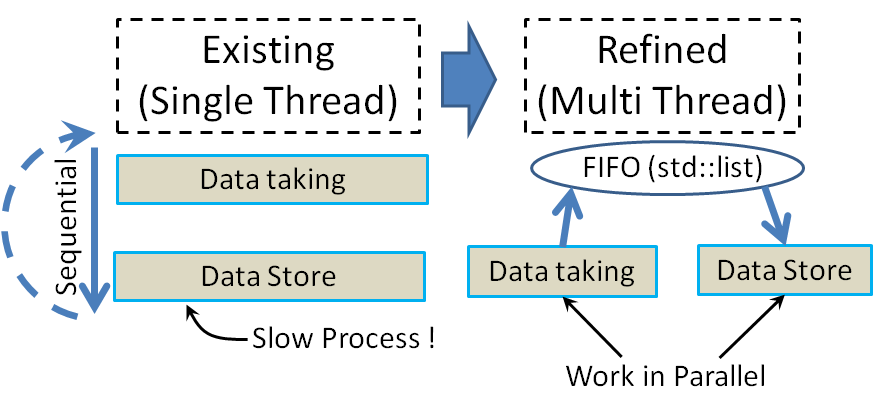}
\caption{The Schema of Multi Thread Processing.}
\label{fig:method_mt}
\end{figure}

\noindent
``Multi Thread Processing" is the other one of the refine approach in software region. 
The schema of this refine is shown in Fig \ref{fig:method_mt}.

In existing system, software working on single thread processing. 
In the case, data taking job this job is data taking from SEABAS and data store job are sequential. 
So next data taking job have to follow previous data store job. 
In result, DAQ whole efficiency is reduced. 

Therefore, we refined this software structure by using multi thread processing. 
To implement multi thread, we use WIN32API\cite{winapi} and Posix Thread\cite{posix}.

Refined software's structure consists of 2 threads, one is data taking job thread, and the other is data store job thread. 
These threads are working in parallel, so data taking don't have to wait the other job. 
To pass data between 2 threads, use ``First In, First Out (FIFO)" buffer based on std::list\cite{cpp}. 
This refine make it possible to take data at the maximum speed. 
From measured value, data transfer rate of refined software is three times as speedy as that of existing one. 

\clearpage

\subsection{Trigger rate control (TRC)}

\begin{figure}[H]
\centering
\includegraphics[width=14cm]{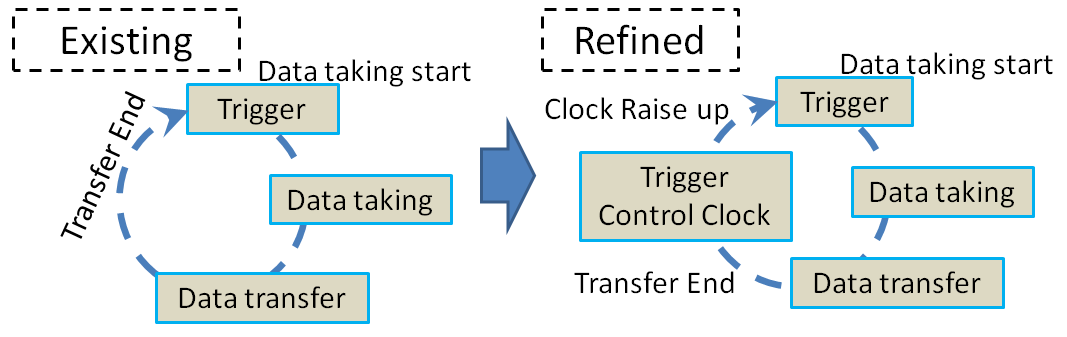}
\caption{The Schema of Trigger rate control function.}
\label{fig:method_trc}
\end{figure}

\noindent
``Trigger rate control" is the refine approach in firmware region. 
The schema of this refine is shown in Fig \ref{fig:method_trc}. 

In existing system, frame data from sensor is taken one after another, without rate control. 
Because of this, sometime frame rate stability is broken when data transfer is delaying. 
To refine this problem, the flow control function is required. 

Therefore, we implemented some low frequency clock as ``Trigger Control Clock (TCC)" to firmware. 
Role of this clock is rate control. 

The flow of the data taking sequence with trigger rate control is this way. 
\begin{enumerate}
 \item When TCC raise up while waiting trigger, start data taking. 
 \item While data taking and transfer, issuing of next trigger is postponed. 
 \item After the end of data transfer, DAQ system will return to wait next trigger state. 
 \item Repeat 1-3 until all frames are taken. 
\end{enumerate}

When TCC is set lower than maximum data transfer rate, 
frame rate will be stabilized in fixed (synchronized to TCC) rate. 

\clearpage

\section{Results}

\subsection{New DAQ Software}

\begin{wrapfigure}{r}{6cm}
\centering
\includegraphics[height=4.5cm]{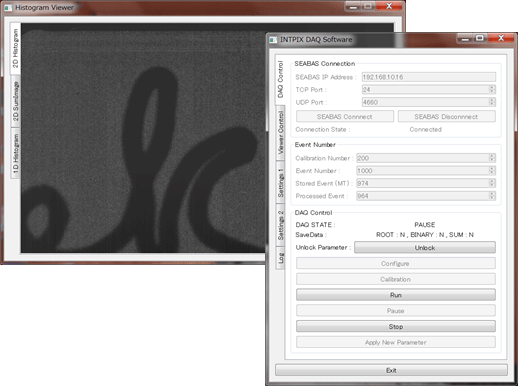}
\caption{The working image of new DAQ software.}
\label{fig:newdaq}
\end{wrapfigure}

We developed the new DAQ software refined from previous software. 
Introduced function, ``Abstraction and Hierarchization" and ``Multi Thread", was implemented.

This software is using some libraries, Qt 5.4, OpenCV2.4.11 and picojson.
Recent version of Windows (Vista, 7, 8, 8.1, both 32 bit and 64 bit) are supported.
Linux will be supported soon.

This software's GUI was completely re-developed based on Qt library. 
This GUI is compatible with previous GUI. 
Thus function's codes can be transplanted each other between new GUI and old one. 

New Software's working image is shown in Fig \ref{fig:newdaq}. 

\subsection{X-ray Imaging}

\begin{figure}[H]
\begin{tabular}{c}
\begin{minipage}{0.55\hsize}
\centering
\includegraphics[width=7cm]{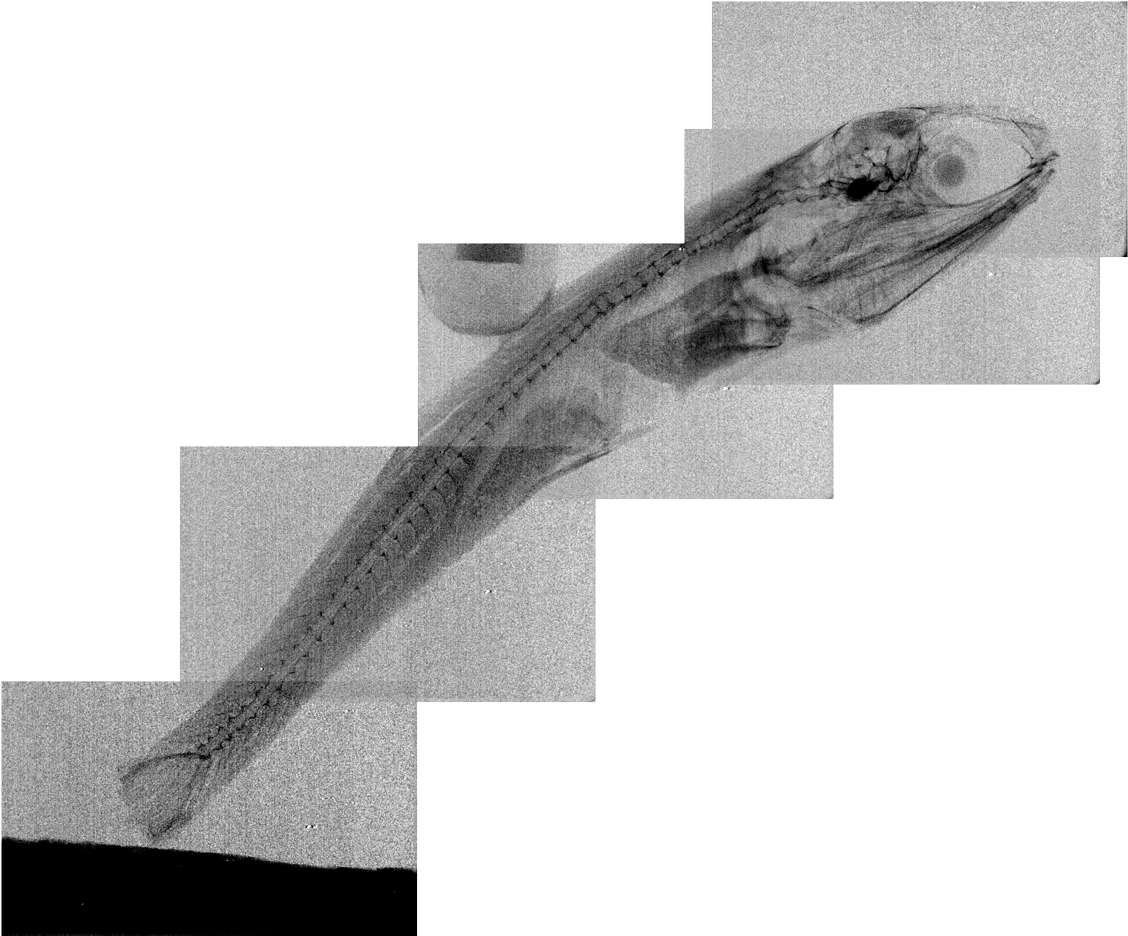}
\caption{The X-ray image of dried anchovy. (combined 5 shifted position's image data.)}
\label{fig:fish}
\end{minipage}
\begin{minipage}{0.45\hsize}
\centering
\includegraphics[width=6cm]{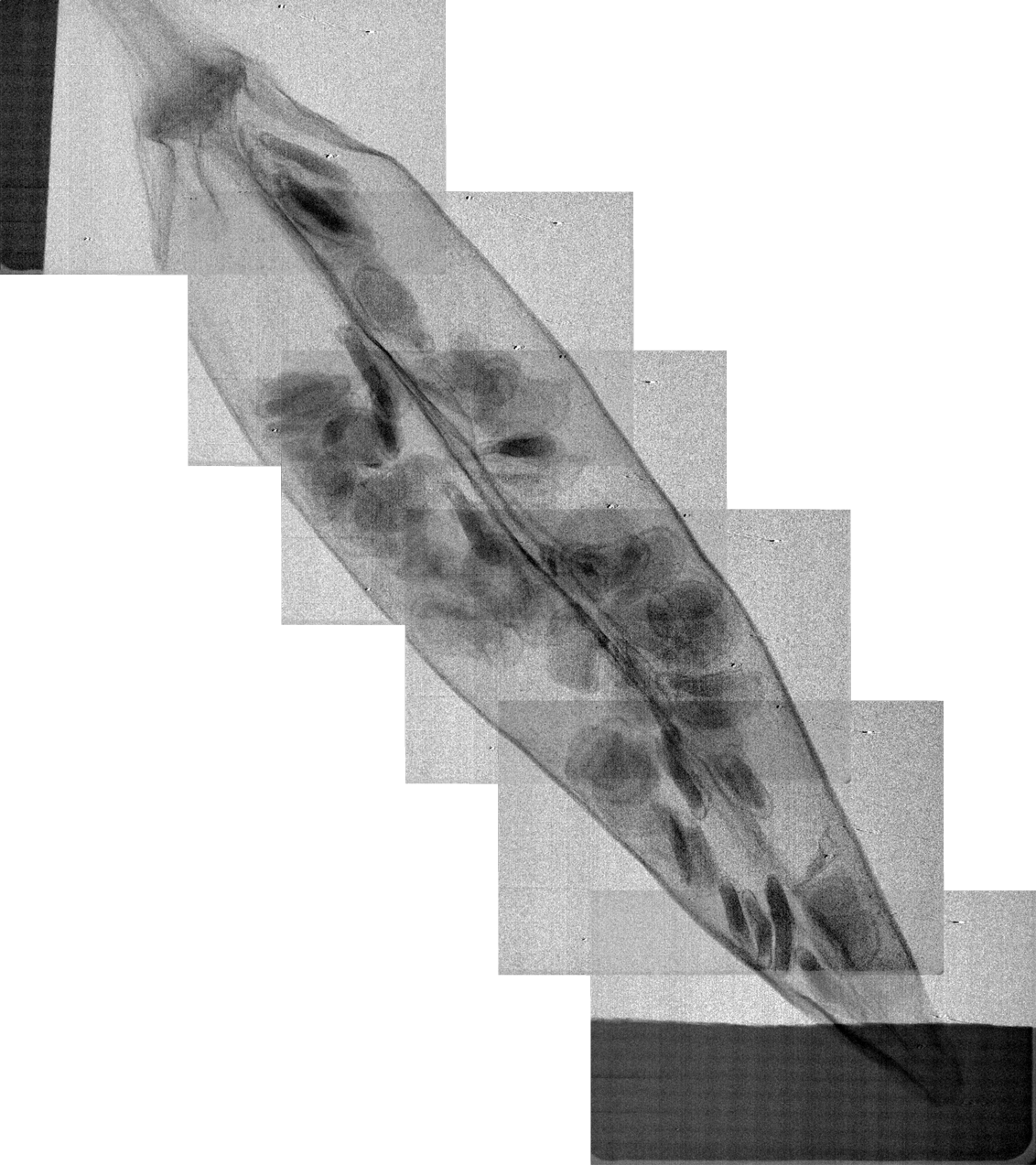}
\caption{The X-ray image of red pepper. (combined 6 shifted position's image data.)}
\label{fig:pepper}
\end{minipage}
\end{tabular}
\end{figure}

This is results of X-ray imaging data taken by new DAQ. 
The Specification of DAQ PC is shown in Table \ref{table:specPC}. 
Fig \ref{fig:fish} is X-ray imaging data of the dried anchovy. 
And Fig \ref{fig:pepper} is X-ray imaging data of the red pepper. 
We can see clearly the structure of sample. 
When we take this, frame rate is 65Hz, this is almost 94\% of estimated maximum. 
And we can confirm new DAQ can take data correctly. 

\begin{table}[H]
  \centering
  \begin{tabular}{|c|c|}
    \hline
    Element & Value \\
    \hline
    \hline
    Model & Lenovo X1 Carbon Gen 3 \\
    \hline
    OS & Windows 7 pro 64 bit \\
    \hline
    CPU & Intel Core i7 5500U 2.40Ghz \\
    \hline
    Primary Disk (OS \& Software) & Intel SCKJF180A5L 180GB M.2(SATA6Gbps) SSD \\
    \hline
    Storage Disk (Data Store) & I-O DATA HDPC-UT2.0D (USB3.0)  \\
    \hline
    Ethernet & Intel Ethernet Connection I218-V \\
    \hline
  \end{tabular}
  \caption{Specification of DAQ PC used for X-ray imaging and TRC test}
  \label{table:specPC}
\end{table}

\subsection{Trigger rate control Test}

We tested trigger rate control function. 
DAQ PC is same PC as X-ray imaging.

\begin{figure}[H]
\centering
\includegraphics[height=10cm]{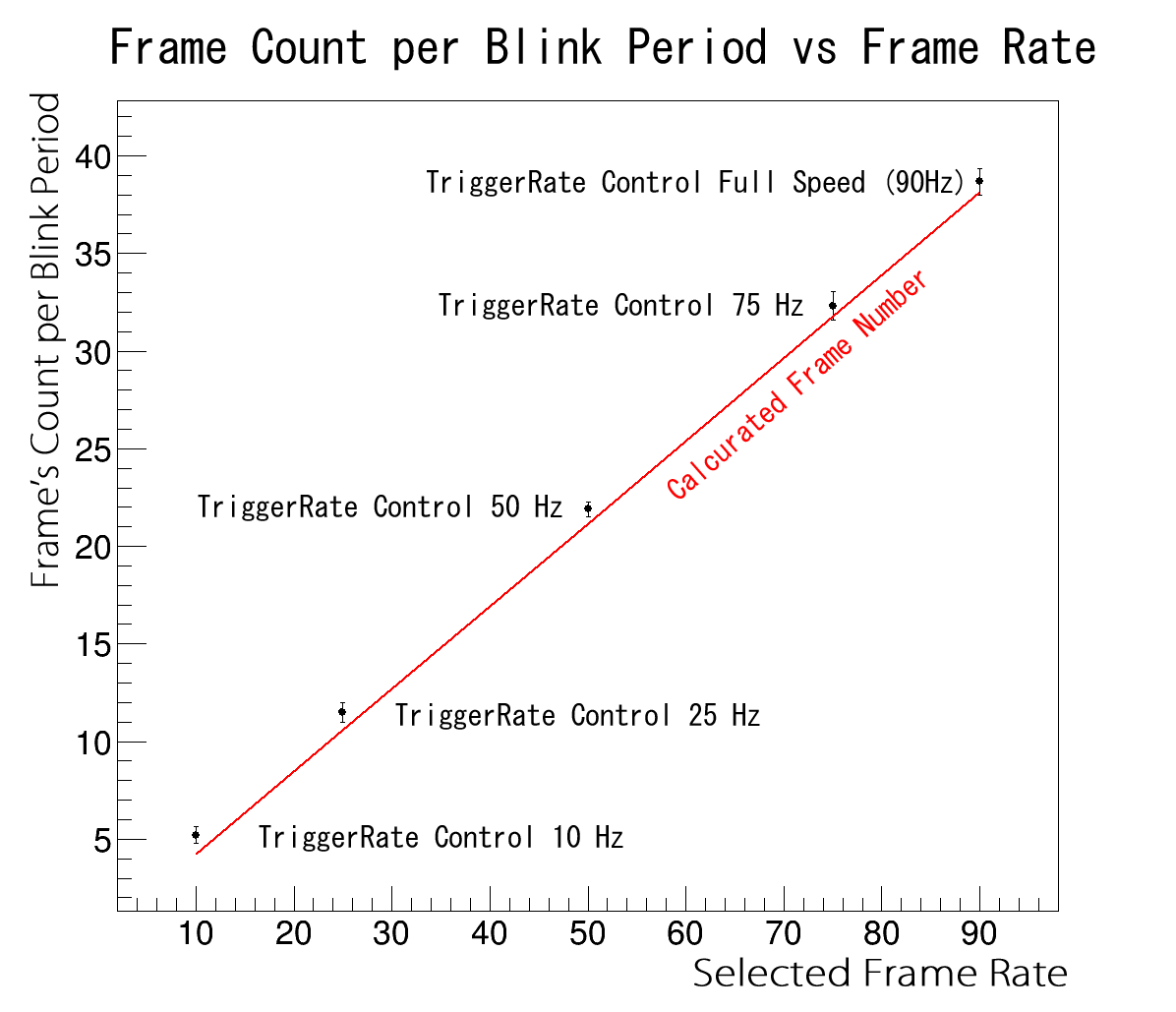}
\caption{The graph of Frame's Period vs Blink Frequency.}
\label{fig:graph1}
\end{figure}

The contents of the test is take a movie of blinking LED (2.4Hz, duty 56\%), and count frames in one blink period. 
When taking data from sensor, set TRC function enable with 4 preset TCC clocks (10, 25, 50, 75 Hz, and Full Speed is noncontrolled). 
Frame's count is related to the real frame rate. 
If TRC works correctly, frame's count will change depending on selected clock. 
Fig \ref{fig:graph1} is result of the test for check relation between selected clock and real frame rate. 
Figure's horizontal axis is selected clock rate, vertical axis is frame's count. 

We can see good relation between frame's count in one blink period and selected frame rate. 

\section{Conclusions}

\begin{itemize}
 \item Developed DAQ firmware \& software for high-speed \& stable readout.
 \item Applied Abstraction and Hierarchization for DAQ software.
 \item Implemented Multi Thread Processing for DAQ software.
 \item New DAQ can high-speed data taking. ($\sim$ 90Hz, 94\% of maximum in X-ray Imaging)
 \item Implemented Trigger rate control for new DAQ, and this function seems work correctly.
 \item We confirmed new DAQ can take data correctly.
\end{itemize}

\Acknowledgements

This work on the SOIPIX group research activity. 
(\url{http://rd.kek.jp/project/soi/})

\end{document}